\documentclass[aps,preprint,showpacs,groupedaddress,floatfix]{revtex4}
\usepackage{graphicx}
\usepackage{dcolumn}
\usepackage{bm}

\begin{document}
\title{A microscopic estimate of the nuclear matter compressibility
and symmetry energy in relativistic mean-field models}
\author{D. Vretenar}
\author{T. Nik\v si\' c}
\affiliation{
Physics Department, Faculty of Science, University of
Zagreb, Croatia, and \\
Physik-Department der Technischen Universit\"at M\"unchen,
D-85748 Garching, Germany}

\author{P. Ring}
\affiliation{Physik-Department der Technischen Universit\"at M\"unchen,
D-85748 Garching, Germany}
\vspace{1cm}
\date{\today}

\begin{abstract}
The relativistic mean-field plus RPA calculations,
based on effective Lagrangians with density-dependent meson-nucleon
vertex functions, are employed in a microscopic analysis
of the nuclear matter compressibility
and symmetry energy. We compute the isoscalar monopole and the
isovector dipole response of $^{208}$Pb, as well as the differences
between the neutron and proton radii for $^{208}$Pb and several Sn isotopes.
The comparison of the calculated excitation energies with the
experimental data on the giant monopole resonance in $^{208}$Pb,
restricts the nuclear matter compression
modulus of structure models based on the relativistic
mean-field approximation to $K_{\rm nm}\approx 250 - 270$ MeV.
The isovector giant dipole resonance in $^{208}$Pb, and the
available data on differences between neutron and proton radii,
limit the range of the nuclear matter symmetry energy at
saturation (volume asymmetry) to 32 MeV $\leq a_4 \leq$ 36 MeV.
\end{abstract}

\maketitle

%
%
\section{\label{secI}Introduction}

Basic properties of nuclear ground states, excitation energies
of giant monopole resonances, the structure of neutron stars, the dynamics
of heavy-ion collisions and of supernovae explosions, depend
on the nuclear matter compressibility. The
nuclear matter compression modulus $K_{\rm nm}$ is defined as
\begin{equation}
K_{\rm nm} = k^2_f {{d^2 E/A}\over {d k^2_f}} \left|_{k_{f_0}}\right. ,
\end{equation}
where $E/A$ is the binding energy per nucleon, $k_{f}$ is the Fermi momentum,
and $k_{f_0}$ is the equilibrium Fermi momentum. The value of $K_{\rm nm}$
cannot be measured directly. In principle it can be extracted
from the experimental energies of isoscalar monopole vibrations
(giant monopole resonances GMR) in nuclei. Semi-empirical macroscopic
leptodermous expansions, as well as microscopic calculations, have been
employed in the analysis of available data on isoscalar GMR. Although
macroscopic expansions, analogous to the liquid drop mass formula,
in principle provide "model independent" estimates of $K_{\rm nm}$,
in reality they do not constrain its value to better than 50\%.

A more reliable approach to the determination of $K_{\rm nm}$ is based
on microscopic calculations of GMR excitation energies.
Self-consistent mean-field calculations of nuclear ground state properties
are performed by using effective interactions with different values of
$K_{\rm nm}$. Interactions that differ in their prediction
of the nuclear matter compressibility,
but otherwise reproduce experimental data
on ground-state properties reasonably well,
are then used to calculate GMR in the random phase approximation
or the time-dependent framework. A fully
self-consistent calculation of both ground-state properties and
GMR excitation energies restricts the range of possible
values for $K_{\rm nm}$.
The correct value of $K_{\rm nm}$ should then be
given by that interaction which reproduces the excitation
energies of GMR in finite nuclei. It has been pointed out,
however, that, since $K_{\rm nm}$ determines
bulk properties of nuclei and, on the other hand,
the GMR excitation energies depend also on the surface compressibility,
the determination of the nuclear matter compressibility relies more on
a good measurement and microscopic calculations of GMR in a
single heavy nucleus such as $^{208}$Pb, rather than on
the systematics over the whole periodic
table~\cite{BBD.95a,Bla.80}. It has also been
emphasized that the determination of a static quantity, $K_{\rm nm}$, from
dynamical properties, i.e. from GMR energies, is potentially
ambiguous. Various dynamical effects, as for instance the
coupling of single particle
and collective degrees of freedom, could modify, though
not much, the deduced value of the nuclear matter compression modulus.
On the other hand, it has been shown that $K_{\rm nm}$ cannot be
extracted from static properties, i.e. masses and charge distributions,
alone~\cite{Far.97}.

In this work we address a different source of ambiguity, which has become
apparent only recently. Modern
non-relativistic Hartree-Fock plus random phase approximation
(RPA) calculations, using both
Skyrme and Gogny effective interactions, indicate that the value of
$K_{\rm nm}$ should be in the range 210-220 MeV~\cite{BBD.95a,Bla.80}.
In Ref.~\cite{Far.97} it has been shown that even generalized Skyrme
forces, with both density- and momentum-dependent terms, can only reproduce
the measured breathing mode energies for values of $K_{\rm nm}$ in the
interval $215\pm 15$ MeV. A comparison of the most recent data on the
$E0$ strength distributions in $^{90}$Zr, $^{116}$Sn, $^{114}$Sm and
$^{208}$Pb, with microscopic calculations based on Gogny effective interactions
by Blaizot {\it et al}.~\cite{BBD.95a}, has put the value of $K_{\rm nm}$
at $231\pm 5$ MeV~\cite{YCL.99}.
In relativistic mean-field (RMF) models on the other hand, results of both
time-dependent and RPA calculations suggest that
empirical GMR energies are best reproduced by an effective
force with $K_{\rm nm}\approx 250 - 270$ MeV~\cite{Vre.97,Vre.99,VWR.00,MGW.01}.
Twenty percent, of course, represents a rather
large difference. The origin of this
discrepancy is at present not understood.

It appears, however, that not all results of the relativistic models
are in agreement either. By using the same type of non-linear
RMF Lagrangians with scalar self-interactions as in
Refs. \cite{Vre.97,Vre.99,VWR.00,MGW.01}, in a series of recent
papers \cite{Pie.62,Pie.64,Pie.66} Piekarewicz has performed fully
consistent relativistic RPA calculations of nuclear compressional modes.
Basically the results can be summarized as follows. (1) The best description
of both compressional modes -- isoscalar monopole and dipole -- is
obtained by using a ``soft" effective interaction with $K_{\rm nm} = 224$ MeV.
(2) Effective interactions with compression moduli
well above $K_{\rm nm} \approx 200$ MeV are not consistent with experimental
data. (3) The compression modulus determined from the
empirical excitation energies of the GMR depends on the nuclear matter
symmetry energy, i.e. models with a lower symmetry energy at saturation
density reproduce the GMR in $^{208}$Pb by using a lower value of $K_{\rm nm}$.
(4) The variance between the values of $K_{\rm nm}$
determined from non-relativistic
and relativistic mean-field plus RPA calculations of GMR excitation energies
can be attributed in part to the differences in the nuclear matter symmetry
energy predicted by non-relativistic and relativistic models.
In particular, in Ref. \cite{Pie.66} it has been pointed out that, when the
symmetry energy of the RMF models is softened  to simulate the symmetry
energy of Skyrme interactions, a lower value of $K_{\rm nm}$, consistent
with the ones used in non-relativistic models, is required to
reproduce the energy of the GMR in $^{208}$Pb.

Our own results strongly disagree with most of the conclusions of
Refs. \cite{Pie.62,Pie.64,Pie.66}. In addition to the results
obtained in the RMF+RPA framework based on non-linear Lagrangians
with scalar self-interactions \cite{Vre.97,Vre.99,VWR.00,MGW.01},
in Ref. \cite{NVR.02} we have carried out calculations of the
isoscalar monopole, isovector dipole and isoscalar quadrupole
response of $^{208}$Pb, in the fully self-consistent relativistic
RPA framework based on effective interactions with a
phenomenological density dependence adjusted to nuclear matter and
ground-state properties of spherical nuclei. The analysis of the
isoscalar monopole response with density dependent coupling
constants has shown that only interactions with the nuclear matter
compression modulus in the range $K_{\rm nm}\approx 260 - 270$
MeV, reproduce the experimental excitation energy of the GMR in
$^{208}$Pb. In addition, the comparison with the experimental
excitation energy of the isovector dipole resonance has
constrained the volume asymmetry to the interval 34 MeV $\leq a_4
\leq$ 36 MeV.

Even though we cannot explain the difference between the
nuclear matter compressibility predicted by non-relativistic and
relativistic mean-field plus RPA calculations, at least we can clarify
the apparent inconsistency of the results obtained in the
relativistic framework.
The purpose of this work is twofold. By using the RMF+RPA
based on effective interactions with density-dependent meson-nucleon
couplings, we will show: (1) contrary to the claims made in Ref. \cite{Pie.66},
our knowledge of the symmetry energy at low density is not that poor, and
the volume asymmetry cannot be lowered to the range of values
$a_4 \leq 30$ MeV, for which relativistic models with
$K_{\rm nm} \leq 230$ MeV  would reproduce the excitation energy
of the GMR in $^{208}$Pb and, (2) relativistic mean-field effective
interactions adjusted to reproduce ground-state properties of spherical
nuclei (binding energies, charge radii, differences between neutron and
proton radii), cannot reproduce the excitation energies of GMR if
$K_{\rm nm} \leq 250$ MeV. Therefore, we will reinforce our
result that $K_{\rm nm} = 250$ MeV is the lower bound for the
nuclear matter compression modulus in nuclear structure models
based on the relativistic mean-field approximation.

%
\section{\label{secII}Effective interactions with
density-dependent meson-nucleon couplings and the relativistic RPA}

The relativistic random phase approximation (RRPA)
will be used to calculate the isoscalar monopole
and isovector dipole strength distributions in $^{208}$Pb.
The RRPA represents the small amplitude limit of the
time-dependent relativistic mean-field theory.
A self-consistent calculations ensures that the same correlations which
define the ground-state properties, also determine
the behavior of small deviations from the equilibrium.
The same effective Lagrangian generates the Dirac-Hartree
single-particle spectrum and the residual particle-hole
interaction. In Ref.~\cite{Daw.90} it has been
shown that an RRPA calculation, consistent with the mean-field
model in the $no-sea$ approximation, necessitates configuration
spaces that include both particle-hole pairs and pairs formed
from occupied states and negative-energy states. The contributions
from configurations built from occupied positive-energy states and
negative-energy states are essential for current conservation and
the decoupling of the spurious state. In addition, configurations
which include negative-energy states give an important contribution
to the collectivity of excited states.
In two recent studies~\cite{MGW.01,Rin.01}
we have shown that a fully consistent inclusion of the Dirac sea of
negative energy states in the RRPA is crucial for a quantitative
comparison with the experimental excitation energies of isoscalar
giant resonances.

The second requisite for a successful application of the RRPA in the
description of dynamical properties of nuclei is the use of
effective Lagrangians with non-linear meson self-interactions,
or Lagrangians characterized by
density-dependent meson-nucleon vertex functions. Even though several
RRPA implementations have been available for almost twenty years,
techniques which enable the inclusion
of non-linear meson interaction terms in the RRPA
have been developed only recently \cite{MGT.97,VWR.00,Pie.62}.
In Ref. \cite{NVR.02} the RRPA matrix equations have been derived
for an effective Lagrangian with density-dependent meson-nucleon
couplings.

Already in Ref. \cite{Vre.97} we used Lagrangians with
non-linear meson self-interactions in time-dependent RMF calculations
of monopole oscillations of spherical nuclei.
The energies of the GMR were determined from the Fourier power spectra
of the time-dependent isoscalar monopole moments $< r^2 > (t)$.
It has been shown that the GMR in heavy nuclei, as well as the
empirical excitation energy curve $E_x \approx 80~A^{-1/3}$ MeV, are
best reproduced by an effective
force with $K_{\rm nm}\approx 250 - 270$ MeV. This result has been
confirmed by the RRPA calculation of Ref. \cite{MGW.01}. In particular,
the best results have been obtained with the NL3 effective
interaction~\cite{LKR.97} ($K_{\rm nm} = 272$ MeV).
Many calculations of ground-state properties and
excited states, performed by different groups, have shown
that NL3 is the best non-linear relativistic
effective interaction so far, both for nuclei at and away from the
line of $\beta $-stability. On the other hand, by using
the same type of effective Lagrangians with non-linear isoscalar-scalar
interaction terms, in the recent RRPA analysis of nuclear compressional modes
of Refs.~\cite{Pie.62,Pie.64} it has been
suggested that models of nuclear structure having
$K_{\rm nm}$ well above $\approx 200$ MeV are likely to be in conflict
with experiment.

Even though models with isoscalar-scalar meson self-interactions have been
used in most applications of RMF to nuclear structure
in the last fifteen years, they
present well known limitations. In the isovector
channel, for instance, these interactions are characterized by large
values of the symmetry energy at saturation -- 37.9 MeV for NL3 -- as
compared with the empirical value $a_4 = 30\pm 4$ MeV.
This is because the isovector channel of these
effective forces is parameterized by a single constant: the $\rho$-meson
nucleon coupling $g_\rho$. With a single parameter in the isovector channel,
it is not possible to reproduce simultaneously the empirical value
of $a_4$ and the masses of $N \neq Z$ nuclei. On the other hand,
extensions of the RMF model that include additional
interaction terms in the isoscalar and/or the isovector channels
were not very successful, at least in nuclear structure applications.
The reason is that the empirical data set of bulk and single-particle
properties of finite nuclei can only constrain six or seven
parameters in the general expansion of an effective
Lagrangian~\cite{FS.00}. Adding more interaction terms does not improve the
description of finite nuclei. Rather, their coupling
parameters and even their forms cannot be accurately determined.

Models based on an effective hadron field theory with medium dependent
meson-nucleon vertices~\cite{FLW.95,TW.99,HKL.01}
present a very successful alternative to the use of
nonlinear self-interactions. Such an approach retains the basic
structure of the relativistic mean-field framework,
but could be more directly related to the
underlying microscopic description of nuclear interactions.
In Ref.~\cite{NVFR.02} we have extended the
relativistic Hartree-Bogoliubov (RHB) model~\cite{PVL.97} to include
density dependent meson-nucleon couplings. The effective Lagrangian is
characterized by a phenomenological density dependence of the $\sigma$,
$\omega$ and $\rho$ meson-nucleon vertex functions, adjusted to properties
of nuclear matter and finite nuclei. It has been shown that, in comparison
with standard RMF effective interactions with nonlinear meson-exchange
terms, the density-dependent meson-nucleon couplings significantly
improve the description of symmetric and asymmetric nuclear matter,
and of ground-state properties of $N\neq Z$ nuclei.

The RRPA with density-dependent meson-nucleon couplings has been derived
in Ref.~\cite{NVR.02}. Just as in the static case the single-nucleon Dirac
equation includes the additional rearrangement self-energies that
result from the variation of the vertex functionals with respect
to the nucleon field operators, the explicit
density dependence of the meson-nucleon couplings introduces
rearrangement terms in the residual two-body interaction. Their
contribution is essential for a quantitative description of excited
states. By constructing families of interactions with some given
characteristic (compressibility, symmetry energy, effective mass),
it has been shown how the comparison of the RRPA results on
multipole giant resonances with experimental data, can be used
to constrain the
parameters that characterize the isoscalar and isovector channel
of the density-dependent effective interactions. In particular,
the GMR in $^{208}$Pb requires the compression modulus to be in the
range $K_{\rm nm}\approx 260 - 270$ MeV, and the isovector GDR in
$^{208}$Pb is only reproduced with the volume asymmetry in the
interval 34 MeV $\leq a_4 \leq$ 36 MeV. We have not, however,
analyzed the influence of the symmetry energy on the range
of allowed values of $K_{\rm nm}$. We have also not tried to
correlate the RRPA results for the isovector GDR with data
on neutron radii, although it has been shown in Ref. \cite{NVFR.02}
that RHB calculations with the density-dependent effective
interaction DD-ME1 ($a_4 = 33.1$ MeV) reproduce the available data
on differences between neutron and proton radii for $^{208}$Pb
and several Sn isotopes.

%
\section{\label{secIII}Nuclear matter compressibility
and symmetry energy}

By performing fully consistent RMF plus RRPA calculations
of nuclear ground-state properties and excitation energies of
giant resonances, in this section we will try to correlate
the nuclear matter symmetry energy and the nuclear matter
compression modulus of relativistic mean-field effective
interactions.

In Ref.~\cite{Pie.66} Piekarewicz has used the standard RMF
effective interactions with isoscalar-scalar meson self-interactions,
and with compression moduli in the range $K_{\rm nm}\approx 200 - 300$ MeV,
to compute the distribution of isoscalar monopole strength
in $^{208}$Pb. The main result of his analysis is that, when the
symmetry energy is artificially softened, in an attempt to
simulate the symmetry energy of Skyrme interactions, a lower
value for the compression modulus is obtained, consistent
with the predictions of non-relativistic Hartree-Fock plus RPA
calculations.

As we have already emphasized in the previous section, the RMF
models with isoscalar-scalar meson self-interactions are characterized
by large values of the symmetry energy at saturation {volume asymmetry}.
When adjusting such an effective interaction to properties of nuclear
matter and bulk ground-state properties of finite nuclei (binding
energy, charge radius) it would be clearly desirable to
bring down the value of $a_4$ to its empirical
value $a_4 = 30\pm 4$ MeV. This, however,
is not feasible. The simple reason is that, if $a_4$ is brought
below $\approx 36 - 37$ MeV by simply reducing the single coupling
constant $g_\rho$ in the isovector channel, then it is no longer possible
to reproduce the relative positions of the neutron and proton Fermi
levels in finite nuclei, i.e. the calculated masses of
$N \neq Z$ nuclei display large deviations from the experimental
values. The binding energies are only reproduced
if a density dependence is included in the $\rho$-meson coupling,
or a non-linear $\rho$-meson self-interaction is included in the
model. If the interaction, however, is adjusted to a single nucleus,
as it was done in Ref.~\cite{Pie.66} for $^{208}$Pb, then even the
standard RMF interactions contain enough parameters to obtain almost any
combination of symmetry energy and nuclear matter compressibility.

In the present analysis we have used effective
interactions with density-dependent meson-nucleon vertex functions.
For the density dependence of the meson-nucleon couplings we adopt the
functionals used in Refs.\cite{TW.99,HKL.01,NVFR.02}.
The coupling of the $\sigma$-meson and $\omega$-meson to the nucleon
field reads
\begin{equation}
g_i(\rho) = g_i(\rho_{\rm sat}) f_i(x)\quad {\rm for}\quad i=\sigma, \omega\;,
\label{coupl}
\end{equation}
where
\begin{equation}
f_i(x) = a_i \frac{1 + b_i(x+d_i)^2}{1 + c_i(x+d_i)^2}
\label{func}
\end{equation}
is a function of $x = \rho /\rho_{\rm sat}$, and $\rho_{\rm sat}$ denotes
the baryon density at saturation in symmetric nuclear matter.
The eight real parameters
in (\ref{func}) are not independent. The five constraints $f_i(1)=1$,
$f_\sigma^{\prime\prime}(1) = f_\omega^{\prime\prime}(1)$, and
$f_i^{\prime\prime}(0)=0$, reduce the number of independent parameters
to three. Three additional parameters in the isoscalar channel are:
$g_\sigma(\rho_{\rm sat})$, $g_\omega(\rho_{\rm sat})$, and
$m_\sigma$ - the mass of the phenomenological sigma-meson.
For the $\rho$-meson coupling the functional
form of the density dependence is suggested by Dirac-Brueckner
calculations of asymmetric nuclear matter~\cite{JL.98}
\begin{equation}
\label{drho}
g_{\rho}(\rho) = g_{\rho}(\rho_{\rm sat})~{\rm exp}
\left [-a_{\rho} (x-1)\right ]\;.
\label{grho}
\end{equation}
The isovector channel is parameterized by $g_{\rho}(\rho_{\rm sat})$ and
$a_{\rho}$. Usually the free values are used
for the masses of the $\omega$ and $\rho$ mesons:
$m_\omega = 783$ MeV and
$m_\rho = 763$ MeV. In principle one could also consider the
density dependence of the meson masses. However, since the effective
meson-nucleon coupling in nuclear matter
is determined by the ratio $g/m$, the choice of a
phenomenological density dependence of the couplings makes an
explicit density dependence of the masses redundant.

Obviously, the framework of density-dependent interactions is more
general than the standard RMF models, and it encloses models
with non-linear meson self-interactions in the isoscalar-scalar,
isoscalar-vector, and isovector-vector channels.
The eight independent parameters, seven coupling parameters and
the mass of the $\sigma$-meson, are adjusted to reproduce the
properties of symmetric and asymmetric nuclear matter,
binding energies and charge radii of spherical nuclei. For the
open-shell nuclei pairing correlations are
treated in the BCS approximation with empirical pairing gaps
(five-point formula).

In order to investigate possible correlations between the nuclear matter
symmetry energy and the compression modulus, we have constructed three
families of interactions, with $K_{\rm nm} =$ 230, 250, and 270 MeV,
respectively. For each value of $K_{\rm nm}$ we have adjusted five
interactions with $a_4 =$ 30, 32, 34, 36 and 38 MeV, respectively.
These interactions have been fitted to properties of nuclear matter
(the binding energy $E/A = 16$ MeV  (5\%), 
the saturation density $\rho_{\rm sat} = 0.153$ fm$^{-3}$  (5\%),
the compression modulus $K_{\rm nm}$ (0.1\%), and the volume asymmetry $a_4$ (0.1\%)),
and to the binding energies (0.1\%) and charge radii (0.2\%) of ten spherical nuclei:
$^{16}$O, $^{40}$Ca, $^{90}$Zr, $^{112,116,124,132}$Sn, and $^{204,208,214}$Pb,
as well as to the differences between neutron and proton radii (10\%) for 
the nuclei $^{116}$Sn, $^{124}$Sn and $^{208}$Pb. The values in
parentheses correspond to the error bars used in the fitting procedure.
Note that, in order to fix an interaction with particular values of 
$K_{\rm nm}$ and $a_4$, a very small error bar of only 0.1\% is
used for these two quantities.
For each family of interactions with a given $K_{\rm nm}$ and for each
$a_4$, in Fig. \ref{figA} we plot the differences, expressed as a
percentage, between the calculated
and experimental binding energies~\cite{AW.95}.
The corresponding deviations of charge radii are shown in Fig. \ref{figB}.
The results are rather good. Most deviations of the binding energies are
$\leq$ 0.2\%, and the differences between the calculated and experimental
charge radii~\cite{CHARGE} are $\leq$ 0.3\%, except for $^{40}$Ca. It should be
emphasized that, by using the standard RMF
models with only isoscalar-scalar meson self-interactions, it is simply
not possible to construct a set of interactions with this
span of values of $K_{\rm nm}$ and $a_4$ and the same quality of
deviations of masses and charge radii from experimental values.

For the three sets of interactions with $K_{\rm nm} =$ 230, 250, and 270 MeV,
in Fig. \ref{figC} we display the corresponding
nuclear matter symmetry energy curves for each choice of the
volume asymmetry $a_4$. The symmetry energy can be parameterized
\begin{equation}
S_2(\rho) = a_4 + \frac{p_0}{\rho_{\rm sat}^2}~(\rho - \rho_{\rm sat}) +
\frac{\Delta K_0}{18 \rho_{\rm sat}^2}~ (\rho - \rho_{\rm sat})^2 + \cdots
\label{S2}
\end{equation}
The parameter $p_0$ defines the linear density dependence
of the symmetry energy, and $\Delta K_0$ is the correction to the
incompressibility.

In finite nuclei, among other quantities,
the symmetry energy directly determines the differences
between the neutron and the proton radii.
In a recent study of neutron
radii in non-relativistic and covariant mean-field models~\cite{Fur.01},
the linear correlation between the neutron skin
and the symmetry energy has been analyzed.
In particular, the analysis has shown that there is a
very strong linear correlation between the neutron skin thickness in
$^{208}$Pb and the individual parameters that determine the
symmetry energy $S_2(\rho)$: $a_4$, $p_0$ and $\Delta K_0$. The empirical
value of $r_n - r_p$ in $^{208}$Pb ($0.20 \pm 0.04$ fm from proton
scattering data~\cite{SH.94}, and $0.19 \pm 0.09$ fm from the alpha scattering
excitation of the isovector giant dipole resonance~\cite{Kra.94}) places the
following constraints on the values of the parameters of the symmetry
energy: $a_4 \approx 30-34$ MeV, 2 Mev/fm$^3 \leq p_0 \leq$ 4 Mev/fm$^3$,
and $-200$ MeV $\leq \Delta K_0 \leq -50$ MeV.

For the three sets of interactions with $K_{\rm nm} =$ 230, 250, and 270 MeV,
in the two lower panels of Fig. \ref{figD} we plot the coefficients
$p_0$ and $\Delta K_0$ as functions of the volume asymmetry $a_4$.
As we have shown in Ref.~\cite{NVR.02}, in order to reproduce
the bulk properties of spherical
nuclei, larger values of $a_4$ necessitate an increase of $p_0$. It is
important to note that only
in the interval 32 MeV $\leq a_4 \leq$ 36 MeV, both $p_0$ and $\Delta K_0$
are found within the bounds determined by the value of $r_n - r_p$ in
$^{208}$Pb. The increase of
$p_0$ with $a_4$ implies a transition from a parabolic to an
almost linear density dependence of $S_2$ in the density region
$\rho \leq 0.2$ fm$^{-3}$ (see Fig. \ref{figC}).
This means, in particular, that the increase
of the asymmetry energy at saturation point will produce an
effective decrease of $S_2$ below $\rho \approx 0.1$ fm$^{-3}$.
In Refs. \cite{Rei.99,NVR.02} it has been shown that, as a result
of the increase of $p_0$ with $a_4$,
the excitation energy of the isovector GDR decreases with
increasing $S_2(\rho_{\rm sat}) \equiv a_4$, because this increase
implies a decrease of $S_2$ at low densities characteristic for
surface modes. This effect is illustrated in the upper panel of
Fig. \ref{figD}, where we plot the calculated excitation energies
of the isovector GDR in $^{208}$Pb as functions of $a_4$, for
each set of interactions with $K_{\rm nm} =$ 230, 250, and 270 MeV.
The calculated centroids of the Lorentzian folded strength distributions are
shown in comparison with the experimental value of $13.3\pm 0.1$~\cite{Rit.93}.
The RRPA excitation energies of the isovector GDR decrease linearly
with $a_4$, and the experimental value favors, for all three families
of interactions, the interval 34 MeV $\leq a_4 \leq$ 36 MeV for the
volume asymmetry.

The results of fully consistent RRPA calculations of the isoscalar
monopole response in $^{208}$Pb are shown in the upper panel of
Fig. \ref{figE}, where we plot the excitation energies of the GMR
for the three families of interactions with $K_{\rm nm} =$ 230,
250, and 270 MeV, respectively, as functions of the volume
asymmetry $a_4$. The shaded area denotes the experimental value:
$E = 14.17\pm 0.28$ MeV~\cite{YCL.99}. For each interaction, in
the lower panel we plot the corresponding result for the
difference of the neutron and proton radius in $^{208}$Pb, in
comparison with, at present, the best experimental value: $0.20
\pm 0.04$ fm from proton scattering data~\cite{SH.94}. The
calculated radii practically do not depend on the compressibility
but, of course, display a strong linear dependence on the volume
asymmetry $a_4$. The comparison with the experimental estimate
limits the possible values of the symmetry energy at saturation to
32 MeV $\leq a_4 \leq$ 36. This interval for $a_4$ is somewhat
wider than the range deduced from the isovector GDR but,
nevertheless, both quantities exclude values $a_4 \leq$ 30 MeV and
$a_4 \geq $ 38 MeV. Coming back to the GMR (upper panel of Fig.
\ref{figE}), we notice that only the set of interactions with
$K_{\rm nm} =$ 270 MeV reproduces the experimental excitation
energy of the GMR for all values of the volume asymmetry $a_4$.
With $K_{\rm nm} =$ 250 MeV, only for the two lowest values of
$a_4$ the RRPA results for the GMR excitation energy are found
within the bounds of the experimental value. $K_{\rm nm} =$ 250
MeV is obviously the lower limit, if not even too low, for the
nuclear matter compression modulus of the relativistic mean-field
effective interactions. This result is completely in accordance
with our previous results obtained with relativistic effective
forces with non-linear meson
self-interactions~\cite{Vre.97,MGW.01}, and with density-dependent
interactions~\cite{NVR.02}. The calculation absolutely excludes
the set of interactions with $K_{\rm nm} =$ 230 MeV, for any value
of $a_4$. The calculated energies are simply too low compared with
the experimental value.

Finally, in Fig. \ref{figF} we plot, for all three families
of interactions, the calculated differences between neutron
and proton radii of Sn isotopes, as functions of the
mass number, in comparison with available experimental
data~\cite{Kra.99}. Similar to the result obtained for
$^{208}$Pb, the calculated values practically do not
depend on the nuclear matter compressibility. While
all the interactions reproduce the isotopic trend of the
experimental data, and we also notice that the error bars 
are rather large, nevertheless the comparison excludes values
$a_4 \leq$ 30 MeV and $a_4 \geq $ 38 MeV.
%
\section{\label{secIV}Summary and conclusions}

In this work we have clarified the apparent inconsistency of the
values of the nuclear matter compression modulus deduced from
relativistic RPA calculations of the giant monopole resonance (GMR) in
$^{208}$Pb. By using standard RMF effective Lagrangians with
non-linear isoscalar scalar meson self-interactions, in a
number of studies (time-dependent RMF, relativistic RPA) we
have shown that the experimental data on GMR in heavy nuclei, as well as the
empirical excitation energy curve $E_x \approx 80~A^{-1/3}$ MeV, are
best reproduced by an effective force with
$K_{\rm nm}\approx 250 - 270$ MeV \cite{Vre.97,Vre.99,VWR.00,MGW.01}.
The best results have been obtained with the well known effective
interaction NL3~\cite{LKR.97} ($K_{\rm nm} = 272$ MeV).
On the other hand, by using the same type of effective Lagrangians
in the fully consistent relativistic RPA calculations of the
isoscalar monopole response, Piekarewicz has concluded that
RMF models with $K_{\rm nm}$ well above $\approx 200$ MeV will
be in conflict with experiment \cite{Pie.62,Pie.64}, and
that the discrepancy between the nuclear matter compressibility
determined from non-relativistic
and relativistic mean-field plus RPA calculations of GMR excitation energies
can be attributed in part to the differences in the nuclear matter symmetry
energy predicted by non-relativistic and relativistic models \cite{Pie.66}.

It is well known, however, that the standard RMF Lagrangians, with
meson self-interactions only in the isoscalar channel, are characterized
by large values of the symmetry energy at saturation
(volume asymmetry) $a_4$. In fact, if the effective interaction in
the isovector channel is parameterized by the single $\rho$-meson nucleon
coupling constant, it is not possible to simultaneously reduce the
value of $a_4$ below $\approx 36 - 37$ MeV, and still reproduce the
experimental binding energies of $N \neq Z$ nuclei. In this
framework it is simply not possible to construct effective
interactions with the volume asymmetry
in the range of empirical values $a_4 = 30\pm 4$ MeV.

In the present analysis we have used effective interactions
with density-dependent meson-nucleon couplings. RMF models
based on an effective hadron field theory with medium dependent
meson-nucleon vertices~\cite{FLW.95}, provide a much better
description of symmetric and asymmetric nuclear matter,
and of ground-state properties of $N\neq Z$ nuclei \cite{TW.99,HKL.01,NVFR.02}.
In order to investigate possible correlations between the volume
asymmetry and the nuclear matter compression modulus, we have
constructed three sets of effective interactions with
$K_{\rm nm} =$ 230, 250, and 270 MeV, and for each value of $K_{\rm nm}$
we have adjusted five
interactions with $a_4 =$ 30, 32, 34, 36 and 38 MeV, respectively.
The interactions have been fitted to reproduce the nuclear matter
saturation properties, as well as the ground states of ten spherical
nuclei.

By employing the fully consistent RMF plus RRPA framework with
density-dependent effective interactions,
we have computed the isoscalar monopole and the isovector dipole
response of $^{208}$Pb, as well as the differences between the
neutron and proton radii for $^{208}$Pb and several Sn isotopes.
The comparison of the calculated excitation energies with the
experimental data on the GMR and isovector GDR in $^{208}$Pb has shown
that: (i) only for $K_{\rm nm} =$ 270 MeV
the RRPA calculation reproduces the experimental
excitation energy of the GMR for all values of the volume
asymmetry $a_4$, (ii) $K_{\rm nm} =$ 250 MeV represents the lower
limit for the nuclear matter compression
modulus of the relativistic mean-field effective interactions,
(iii) the isovector GDR constrains
the volume asymmetry to the interval 34 MeV $\leq a_4 \leq$ 36 MeV.
In comparison with the available experimental data, the
calculated differences between neutron and proton radii indicate
that the volume asymmetry should be in the range
32 MeV $\leq a_4 \leq$ 36 MeV, and reinforce our conclusion that
$a_4$ cannot be lowered to a range of values
for which relativistic models with
$K_{\rm nm} \leq 230$ MeV  would reproduce the excitation energy
of the GMR in $^{208}$Pb.
The disagreement between the nuclear matter compression moduli
predicted by non-relativistic and relativistic mean-field plus
RPA calculations, cannot be explained by the
differences in the volume asymmetry
of the non-relativistic and relativistic mean-field models.

The present analysis has confirmed our
earlier results that the nuclear matter compression
modulus of structure models based on the relativistic
mean-field approximation should be restricted to the interval
$K_{\rm nm}\approx 250 - 270$ MeV, or even slightly higher.

In addition, we have also shown that, for the relativistic
mean-field models, the isovector GDR and the
available data on differences between neutron and proton radii
limit the range of the nuclear matter symmetry energy at
saturation to 32 MeV $\leq a_4 \leq$ 36 MeV. It appears that
the GDR favors the high end of this interval, but we stress the
fact that in the present analysis we have not taken into account
the influence of the effective mass on the
calculated excitation energy of the GDR. This is, however,
an effect which really goes beyond the mean-field approximation.
Rather, more accurate data on neutron radii in heavy nuclei
would provide very useful information of the isovector channel
of the effective RMF interactions. On the other hand, as it
has been shown in Ref.~\cite{BBD.95a}, there is no
correlation between the effective mass and the excitation
energy of the GMR. The
choice of the effective mass, therefore, cannot influence
the nuclear matter compressibility extracted from the GMR.

\bigskip

{\bf ACKNOWLEDGMENTS}

This work has been supported in part by the
Bundesministerium f\"{u}r Bildung und Forschung under the project 06 TM 979
and by the Deutsche Forschungsgemeinschaft.

\newpage
\begin{figure}
\caption{The deviations (in percent) of the theoretical binding
energies of ten spherical nuclei, calculated with the three
families of interactions with $K_{\rm nm} =$ 230, 250, and 270
MeV, from the empirical values~\protect\cite{AW.95}. The legend
relates the different symbols to the volume asymmetries of the
corresponding effective interactions.}%
\label{figA}
\end{figure}

\begin{figure}
\caption{Same as in Fig.~\protect\ref{figA}, but for the charge
radii compared with the experimental
values~\protect\cite{CHARGE}.}%
\label{figB}
\end{figure}

\begin{figure}
\caption{$S_2(\rho)$ coefficient (\protect\ref{S2}) of the
quadratic term of the energy per particle of asymmetric nuclear
matter, calculated with the three families of effective
interactions with $K_{\rm nm} =$ 230, 250, and 270 MeV. The legend
relates the different curves to the volume asymmetries of the
corresponding effective interactions.}%
\label{figC}
\end{figure}

\begin{figure}
\caption{The isovector GDR excitation energy of $^{208}$Pb (upper
left panel),the parameter $p_0$ of the linear density dependence
of the nuclear matter asymmetry energy (lower left), and the
correction to the incompressibility $\Delta K_0$ (lower right), as
functions of the volume asymmetry $a_4$. The shaded area denotes
the experimental isovector GD resonance energy $13.3\pm 0.1$ MeV.
The three sets of symbols correspond to the families of
interactions with $K_{\rm nm} =$ 230, 250, and 270 MeV.}
\label{figD}
\end{figure}

\begin{figure}
\caption{The RRPA excitation energies of the GMR in $^{208}$Pb as
functions of the volume asymmetry $a_4$, calculated for the three
sets of interactions with $K_{\rm nm} =$ 230, 250, and 270 MeV.
The theoretical centroids are shown in comparison with the
experimental excitation energy of the monopole resonance: $E =
14.17\pm 0.28$ MeV \protect\cite{YCL.99}. In the lower panel the
corresponding results for the difference of the neutron and proton
radius in $^{208}$Pb, are plotted in comparison with the
experimental value $0.20 \pm 0.04$ fm \protect\cite{SH.94}.}
\label{figE}
\end{figure}

\begin{figure}
\caption{The calculated differences between neutron and proton
radii of Sn isotopes, calculated with the three sets of
interactions with $K_{\rm nm} =$ 230, 250, and 270 MeV, and with
the values of the volume asymmetry $a_4 =$ 30, 32, 34, 36 and 38
MeV. The theoretical values are compared with experimental data
from Ref.~\protect\cite{Kra.99}.}
\label{figF}
\end{figure}


\bigskip

\begin{references}
\bibitem{BBD.95a}J.P. Blaizot, J.F. Berger, J. Decharg\'e, and M. Girod,
    Nucl. Phys. {\bf A591}, 435 (1995).

\bibitem{Bla.80} J.P. Blaizot, Phys. Rep. {\bf 64}, 171 (1980).

\bibitem{Far.97} M. Farine, J.M. Pearson, and F. Tondeur,
    Nucl. Phys. {\bf A615}, 135 (1997).

\bibitem{YCL.99} D.H. Youngblood, H.L. Clark, and Y.-W. Lui,
        Phys. Rev. Lett. {\bf 82}, 691 (1999).

\bibitem{Vre.97} D. Vretenar, G.A. Lalazissis, R. Behnsch,
      W. P\" oschl, and P. Ring, Nucl. Phys. {\bf A621}, 853 (1997).

\bibitem{Vre.99} D. Vretenar, P. Ring, G.A. Lalazissis, and N. Paar,
          Nucl. Phys. {\bf A649}, 29c (1999).

\bibitem{VWR.00} D. Vretenar, A. Wandelt, and P. Ring,
        Phys. Lett. {\bf B487}, 334 (2000).

\bibitem{MGW.01}  Z.Y. Ma, N. Van Giai, A. Wandelt, D. Vretenar, and P.
    Ring, Nucl. Phys. {\bf A686}, 173 (2001).

\bibitem{Pie.62} J. Piekarewicz, Phys. Rev. C {\bf 62}, 051304(R) (2000).

\bibitem{Pie.64} J. Piekarewicz, Phys. Rev. C {\bf 64}, 024307 (2001).

\bibitem{Pie.66} J. Piekarewicz, Phys. Rev. C {\bf 66}, 034305 (2002).

\bibitem{NVR.02} T. Nik\v si\' c, D. Vretenar, and
        P. Ring,  Phys. Rev. C {\bf 66}, 064302 (2002).

\bibitem{Daw.90} J.F. Dawson and R.J. Furnstahl,
    Phys. Rev. C {\bf 42}, 2009 (1990).

\bibitem{Rin.01}  P. Ring, Zhong-yu Ma, Nguyen Van Giai, D. Vretenar,
     A. Wandelt, and Li-gang Cao,
     Nucl. Phys. {\bf A694}, 249 (2001).

\bibitem{MGT.97} Z.Y. Ma, N. Van Giai, H. Toki, and M. L'Huillier,
                 Phys.Rev. C {\bf 42}, 2385 (1997).

\bibitem{LKR.97} G.A. Lalazissis, J. K\"onig, and P. Ring,
    Phys. Rev. C {\bf 55}, 540 (1997).

\bibitem{FS.00} R.J. Furnstahl and B.D. Serot,
    Nucl. Phys. A {\bf 671}, 447 (2000).

\bibitem{FLW.95} C. Fuchs, H. Lenske, and H.H. Wolter,
        Phys. Rev. C {\bf 52}, 3043 (1995).

\bibitem{TW.99} S. Typel and H.H. Wolter,
     Nucl. Phys.  {\bf A656}, 331 (1999).

\bibitem{HKL.01} F. Hofmann, C.M. Keil, and H. Lenske,
    Phys. Rev. C {\bf 64}, 034314 (2001).

\bibitem{NVFR.02} T. Nik\v si\' c, D. Vretenar, P. Finelli, and
    P. Ring, Phys. Rev. C {\bf 66}, 024306 (2002).

\bibitem{PVL.97} W. P\"oschl, D. Vretenar, G.A. Lalazissis,
        and P. Ring, Phys. Rev. Lett. {\bf 79}, 3841 (1997).

\bibitem{JL.98} F. de Jong and H. Lenske,
     Phys. Rev. C {\bf 57}, 3099 (1998).

\bibitem{AW.95} G. Audi and A. H. Wapstra,
    Nucl. Phys. {\bf A595}, 409 (1995).

\bibitem{CHARGE} E.G. Nadiakov, K.P. Marinova, Yu.P. Gangrsky,
              At. Data Nucl. Data Tables {\bf 56}, 133 (1994).

\bibitem{Fur.01} R.J. Furnstahl, Nucl. Phys. {\bf A706}, 85 (2002).

\bibitem{SH.94} V.E. Starodubsky and N.M. Hintz,
     Phys. Rev. C {\bf 49}, 2118 (1994).

\bibitem{Kra.94} A. Krasznahorkay et al.,
    Nucl. Phys. {\bf A567}, 521 (1994).

\bibitem{Rei.99} P.-G. Reinhard, Nucl. Phys. {\bf A649}, 305c (1999).

\bibitem{Rit.93} J. Ritman et al.,
    Phys. Rev. Lett. {\bf 70}, 533 (1993).

\bibitem{Kra.99} A. Krasznahorkay et al.,
    Phys. Rev. Lett. {\bf 82}, 3216 (1999).

\end{references}
\end{document}